\documentclass[12pt]{article}
\usepackage{amsmath,amssymb,mathtools,amsfonts,epsfig,graphicx,euscript}
 \usepackage{color}
 \usepackage{etoolbox} 
 \usepackage{hyperref}
 \vspace{4cm}
\begin{document}
\begin{titlepage}
\thispagestyle{empty}
\begin{center}
\font\titlerm=cmr10 scaled\magstep4
\font\titlei=cmmi10
scaled\magstep4 \font\titleis=cmmi7 scaled\magstep4 {\Large{\textbf{Electroweak baryogenesis via chiral gravitational waves}
\\}}
\setcounter{footnote}{0}
\vspace{1.5cm}
 \noindent{{
H. Abedi$ ^\dagger$ \footnote{e-mail: hamid\_abedi@sbu.ac.ir}, M. Ahmadvand$ ^\ddagger$  \footnote{e-mail: ahmadvand@shahroodut.ac.ir} and S.S. Gousheh$ ^\dagger$\footnote{e-mail: ss-gousheh@sbu.ac.ir}
}}\\
\vspace{0.2cm}

$ ^\dagger${\it Department of Physics, Shahid Beheshti University, G.C., Evin, Tehran 19839, Iran\\}
$ ^\ddagger${\it Physics Department, Shahrood University of Technology, P.O.Box 3619995161 Shahrood, Iran\\}
\vspace*{.4cm}
\today
\end{center}
\vskip 2em
\setcounter{footnote}{0} 
\begin{abstract}
We propose a new mechanism for electroweak baryogenesis based on gravitational waves generated by helical magnetic fields that are present during a first order electroweak phase transition. We generate a net lepton number through the gravitational chiral anomaly which appears due to the chiral gravitational waves produced by these magnetic fields. The observed value of baryon asymmetry can be obtained in our mechanism within parameter space of scenarios with an inverse cascade evolution for magnetic fields which can also be candidates for large-scale magnetic fields.
\end{abstract}
\end{titlepage}
\section{Introduction}
Cosmological evidence implies the excess of matter over antimatter in the Universe. This asymmetry is characterized by $ \eta _B \equiv n_B /s $, where $ n_B $ is the net baryon number density and $ s $ is the entropy density of the Universe. Based on the Big Bang nucleosynthesis and cosmological abundances of light nuclei, this ratio is determined to be $ \eta _B =(0.84\pm 0.07)\times 10^{-10} $, which is in agreement with CMB observations \cite{Patrignani:2016xqp}. \\
To explain the baryon asymmetry of the Universe each scenario of interest should contain three conditions proposed by Sakharov \cite{Sakharov:1967dj}: 1- baryon number violation 2- C and CP violation, 3- departure from thermal equilibrium. Baryon production scenarios suggested during the Electroweak Phase Transition (EWPT), which is one of cosmological PTs manifestly containing the third condition, are known as EW baryogenesis. In the Standard Model (SM), the first Sakharov condition can be achieved by the triangle anomaly
\begin{equation}
\partial_{\mu}J_B^{\mu}=\partial_{\mu}J_L^{\mu}=\frac{3g^2}{64\pi ^2}\epsilon ^{\mu\nu\alpha\beta}W^a_{\mu\nu}W^a_{\alpha\beta}-\frac{3g'^2}{64\pi ^2}\epsilon ^{\mu\nu\alpha\beta}F_{\mu\nu}F_{\alpha\beta}, 
\end{equation}
 where $ J_B^{\mu} $, $ J_L^{\mu} $ are baryon and lepton currents, respectively, $ W^a_{\mu\nu} $ is the SU(2) field strength and $ F_{\mu\nu} $ is the U(1) field strength. The second term can contribute to the baryon and lepton number violation in the case of helical gauge field \cite{Zadeh:2015oqf}. Moreover, in the SM there is a gravitational chiral anomaly which can lead to a lepton number violation \cite{Alexander:2004us}. This anomaly is given by
 \begin{equation}
 \partial_{\mu}J_L^{\mu}=\frac{N}{32\pi ^2}\epsilon ^{\mu\nu\alpha\beta}R_{\mu\nu\rho\sigma}R^{\rho\sigma}_{\alpha\beta}=\frac{N}{16\pi ^2}R \tilde{R}, 
 \end{equation}
where $ N=3 $ in the SM due to different number of left and right-handed degrees of freedom in the leptonic sector, whereas in beyond the SM it can be less than 3 \cite{Drewes:2013gca}. Also, $ R_{\mu\nu\rho\sigma} $ denotes the curvature tensor of the space-time. The value of quantity $ R \tilde{R} $ does not change from its initial value which is supposedly zero unless chiral components of the metric evolve differently. This can be achieved if there is a CP violating source in the system. Due to this gravitational anomaly, chiral leptons and antileptons can be generated in the processes. Sphalerons act on left-handed leptons and convert them into antiquarks and also act on right-handed antileptons and convert them into quarks. These rival processes do not lead to a net matter asymmetry without a CP violating source. In the case of a strong first-order EWPT, sphaleron-mediated processes are suppressed in the broken phase and the produced matter asymmetry is preserved \cite{Cline:2006ts}. However, the usual scenarios for the EW baryogenesis within the SM cannot account for the observed baryon asymmetry since strongly first-order PT and sufficient CP violation cannot be provided. As a consequence, many beyond the SMs including supersymmetric SMs and SMs with an extended Higgs sector have been proposed to solve this puzzle \cite{Ahmadvand:2013sna}. In these models, due to electroweak symmetry breaking, a first-order PT at which two thermodynamical states are separated through bubble walls is fulfilled. \\
During the history of the Universe, cosmic first-order PTs are important yet from another aspect. That is, they are sources of the Gravitational Wave (GW) radiation which can not only be a powerful probe for the early Universe, but also impact its evolution (see \cite{Kakizaki:2015wua} for GW production at EWPT and \cite{Ahmadvand:2017xrw} for QCD PT with a holographic approach). Three different mechanisms have been proposed for the production of these GWs: the collision of bubbles nucleated during a first-order PT \cite{Kosowsky:1992rz}, sound waves \cite{Hindmarsh:2013xza}, and Magnetohydrodynamic (MHD) turbulence \cite{Kosowsky:2001xp} produced by turbulent fluid and magnetic field in the plasma.\\
In this work, we propose a novel mechanism for baryon production during a first-order EWPT based on the gravitational anomaly. We show that this effect can be as important as other conventional mechanisms proposed for the electroweak baryogenesis. In fact, during a first-order EWPT the GWs produced due to the presence of helical (chiral) magnetic fields are also chiral so that left and right-handed fluctuations of the metric components have different dispersion relations. In addition, the helical component of the magnetic field provides a CP violating source in the model. We demonstrate that this mechanism leads to a non-zero gravitational anomaly and find that for magnetic field values compatible with large-scale magnetic fields observed today, baryon asymmetry relying on sphaleron processes can be explained. Although it is possible to choose a specific model, our mechanism works for any extension of the SM which provides a first-order EWPT and generates chiral GWs. Henceforth, we shall assume that such a strong first order EWPT is provided by an extension of the SM.\\
In Section 2, we express the gravitational anomaly in terms of FRW metric perturbations, and then study the magnetic field generated in a first-order EWPT and derive the energy momentum tensor for such magnetic fields contributing to generation of chiral gravitational waves. Subsequently, we solve the equation of motion for these GWs and calculate the gravitational anomaly term and the lepton number density. Then, the numerical results for baryon asymmetry is presented. Finally, we present our conclusions in Section 3.

\section{The electroweak baryogenesis mechanism}
\subsection{Metric perturbations}
A homogeneous and isotropic universe is described by FRW metric which has no contribution to $ \langle R\tilde{R}\rangle $, where $ \langle\rangle $ denotes the quantum expectation value. The generic perturbed form of this background may be parametrized as 
\begin{equation}
ds^2=a^2\Big\{-(1+2\phi)d\tau ^2+v_i d\tau dx^i+\left[(1+2\varphi)\delta_{ij}+h_{ij}\right]dx^idx^j\Big\},
\end{equation}
where $ a(\tau) $ is the scale factor and $ d\tau =a^{-1}dt $ is the conformal time. Also, $ \phi $, $ \varphi $, $ v_i $, and $ h_{ij} $ are scalar, vector, and tensor perturbations of the metric, respectively. Among these perturbations only $ h_{ij} $ has non-vanishing contribution to $ \langle R\tilde{R}\rangle $. Hence, we consider tensor modes as GW polarization and neglect other fluctuations. Furthermore, we restrict to the transverse and traceless gauge for $ h_{ij} $. In this gauge, one can write $  R\tilde{R} $ as follows \cite{Maleknejad:2014wsa}
\begin{equation}
R\tilde{R}=\frac{-2\epsilon ^{ijk}}{a^4}\Big(\ddot{h}_{jl}\partial _i\dot{h}_{lk}-\partial _m\dot{h}_{jl}\partial ^2_{im}h_{lk}+\partial _l\dot{h}_{jm}\partial ^2_{mi}h_{kl} \Big)+\mathcal{O}(h^4) ,
\end{equation}
where $ \dot{h}_{ij}\equiv \partial _{\tau}h_{ij} $. We assume the GWs propagate in the $ z $ direction. The left and right-handed polarizations can be defined as $ h_{R, L}\equiv (h_{11}\pm i h_{12})/2 $. The term $  R\tilde{R} $ is odd under the parity operation which exchanges the left and right components and can be generated only if these components have different dispersion relations. In the next part, we investigate the helical magnetic field as the source of these GW components.
\subsection{Energy momentum tensor of helical magnetic field }
During a first-order EWPT with high Reynolds number, bubble collisions processes give rise to turbulence and charge separation in the plasma. This process leads to the generation of magnetic fields which can be helical due to chiral anomaly during this era, as pointed out by many investigations \cite{Kahniashvili:2005qi}. In addition, in order to consider the coupling between the turbulent fluid and magnetic fields, MHD effects must be taken into account. On the other hand, these helical magnetic fields can create chiral GWs and as we will see in the following, these birefringent GWs bring about non-zero $ R\tilde{R} $. The equation of motion for the GWs in the radiation-dominated epoch of the Universe is given by \cite{Caprini:2007xq}:
\begin{equation}
\ddot{h}_{R,L}(k, \tau)+2\mathcal{H}\dot{h}_{R,L}(k, \tau)+k^2h_{R,L}(k, \tau)=\frac{a^2(\tau)}{M_p^2}\Pi^{\pm}(k),
\end{equation}
where $ \mathcal{H}=aH $, $ H $ is the Hubble expansion rate, $ M_p =2.44\times 10^{18}~\mathrm{GeV} $ is the reduced Planck mass and $ \Pi^{\pm} $ are the sources for $ h_{R,L} $, respectively. Moreover, $ a(\tau)\sim H_0\tau\sqrt{\Omega _{rad}} $ where $ H_0\simeq 10^{-43}~\mathrm{GeV} $ is the Hubble rate today and $ \Omega _{rad}\simeq 10^{-4} $ is the radiation density parameter today. $ \Pi^{\pm} $ are components of $ \Pi_{ij} $ in the helical basis such that $ \Pi_{ij}(k)=e_{ij}^+\Pi^+(k) +e_{ij}^-\Pi^-(k) $ and $ \Pi_{ij}(k)=(P_{im}P_{jn}-\frac{1}{2}P_{ij}P_{mn})T_{mn}(k) $ is the transverse and traceless part of anisotropic energy-momentum tensor which is given by \cite{Caprini:2003vc}
\begin{equation}
T_{ij}(\mathbf{k})=\frac{1}{2(2\pi)^4}\int d^3p\Big[B_i(\mathbf{p})B^*_j(\mathbf{p}-\mathbf{k})-\frac{1}{2}B_m(\mathbf{p})B^*_m(\mathbf{p}-\mathbf{k})\delta _{ij} \Big], 
\end{equation}
where $ P_{ij}=\delta _{ij}-\hat{k}_i\hat{k}_j  $ is the transverse projector and $ B_i $ is the magnetic field. Considering the interaction of magnetic fields with plasma and back-reaction effects, at small scales the viscous plasma and turbulent decay force the energy spectrum of the magnetic field to be exponentially damped. This damping effect can be taken into account by introducing an UV cutoff, $ k_d $, in the spectrum \cite{Caprini:2003vc}. Assuming that the helical magnetic field in momentum space is a stochastic quantity,  it can be described by a Gaussian profile with the UV cutoff, $ k_d=1/\lambda _d $, where $ \lambda _d $ is the dissipation length in the spectrum. For a stochastic magnetic field, all required quantities are characterized by its two-point correlation function. If primordial magnetic fields are produced by causal mechanisms, the magnetic field correlation length should be shorter than the horizon and its two-point correlation function vanishes on scales larger than the horizon. Therefore, the correlation function has compact support. Moreover, the magnetic field is divergence-free and we expect its correlation function to be square-integrable. Then, according to the Paley-Wiener-Schwartz theorem \cite{Schwartz1952}, its Fourier transform is an analytic function for any value of $ k $. The most general ansatz containing such properties for the magnetic field two-point function in momentum space which respects stochastic homogeneity and isotropy can be considered as \cite{Caprini:2003vc, Durrer:1999bk, Durrer:2003ja}
\begin{equation}\label{bb1}
\langle B_i(\mathbf{k})B_j^*(\mathbf{k}') \rangle =\frac{(2\pi)^3}{2}\delta (\mathbf{k}-\mathbf{k}')[(\delta _{ij}-\hat{k}_i\hat{k}_j)S(k)+i\epsilon _{ijm}\hat{k}_mA(k)]. 
\end{equation}
where $ S(k) $ and $ A(k) $ are the symmetric and helical components of the magnetic field two-point function, respectively, which are also analytic functions and have Taylor expansions. For wavenumbers smaller than $ k_d $, which as we shall show is of the order of $ 10^{-3}~\mathrm{GeV} $, we can model them by a simple power law as $ S(k)=S_0k^{n_S} $ and $ A(k)=A_0k^{n_A} $ \cite{Caprini:2003vc, Durrer:1999bk, Durrer:2003ja}. The requirement of analyticity for the term containing $ \hat{k}_i\hat{k}_j $ implies that $ n_S\geq 2 $. Since $ S(k)\propto \langle |\mathbf{B(\mathbf{k})}\cdot \mathbf{B(\mathbf{-k'})}|\rangle $ and $ A(k)\propto |\langle (\mathbf{k} \times \mathbf{B(\mathbf{k})})\cdot \mathbf{B(\mathbf{-k'})}\rangle | $ in the limit $ \mathbf{k'}\rightarrow \mathbf{k} $, for small wavenumbers we expect $ n_A > n_S $. Hence, for wavenumbers smaller than $ k_d $, we choose the widely used forms $ S(k)=S_0k^{2} $ and $ A(k)=A_0k^{3} $, where the constants $ S_0 $ and $ A_0 $ can be fixed by the magnetic field energy density, $ B^2 $, and the averaged helicity, $ \mathcal{B}^2 $, respectively.\\
To express the tensor source in terms of the magnetic field two-point correlation function, one can parametrize the two-point function of the tensor source as \cite{Caprini:2003vc, Durrer:1999bk}
\begin{equation}
\langle \Pi_{ij}(\mathbf{k})\Pi_{mn}^*(\mathbf{k}') \rangle \equiv \frac{(2\pi)^3}{4}\delta (\mathbf{k}-\mathbf{k}')[\mathcal{M}_{ijmn}f(k)+i\mathcal{A} _{ijmn}g(k)], 
\end{equation}
where
\begin{eqnarray}
\mathcal{M}_{ijmn}&\equiv &P_{im}P_{jn}+P_{in}P_{jm}-P_{ij}P_{mn},\nonumber \\
\mathcal{A}_{ijmn}&\equiv &\frac{\hat{k}_l}{2}(P_{jn}\epsilon_{iml}+P_{im}\epsilon_{jnl}+P_{in}\epsilon_{inl}+P_{jm}\epsilon_{inl}). 
\end{eqnarray}
Moreover, in the helical basis the two-point function can be written as \cite{Caprini:2003vc}
\begin{eqnarray}
\langle \Pi^-(k)\Pi^{-*}(k) \rangle &=&|\Pi^-(k)|^2~=~\frac{1}{3}(f(k)+g(k)),\nonumber \\
\langle \Pi^+(k)\Pi^{+*}(k) \rangle &=& |\Pi^+(k)|^2~=~\frac{1}{3}(f(k)-g(k)).
\end{eqnarray}
Finally, $ f(k) $ and $ g(k) $ are obtained from $ S(k) $ and $ A(k) $ as
\begin{eqnarray}
f(k)&=&\frac{2}{(4\pi)^5}\int d^3p\Big(S(p)S(|\mathbf{k}-\mathbf{p}|)(1+\gamma^{2})(1+\beta^{2})+4A(p)A(|\mathbf{k}-\mathbf{p}|)(\gamma \beta)\Big),\nonumber \\
g(k)&=&\frac{8}{(4\pi)^5}\int d^3p\Big(S(p)A(|\mathbf{k}-\mathbf{p}|)(1+\gamma^{2})\beta\Big), 
\end{eqnarray}
where $ \gamma =\hat{\mathbf{k}}\cdot\hat{\mathbf{p}} $ and $ \beta =\hat{\mathbf{k}}\cdot (\widehat{\mathbf{k-p}}) $. As mentioned above, for magnetic fields produced from causal processes, we can take $ S(k)=S_0k^{2} $ and $ A(k)=A_0k^{3} $ up to the UV cutoff, and then in terms of $ B^2 $ and $ \mathcal{B}^2 $, $ f(k) $ and $ g(k) $ can be expressed as \cite{Caprini:2003vc}
\begin{eqnarray}\label{fg}
f(k)&\simeq &\frac{\lambda ^3}{14} \Big(\frac{B^2}{2\Gamma (\frac{5}{2})} \Big)^2 \Big(\frac{\lambda}{\lambda _d}\Big)^7 -\frac{2\lambda ^3}{27} \Big(\frac{\mathcal{B}^2}{2\Gamma (\frac{7}{2})} \Big)^2 \Big(\frac{\lambda}{\lambda _d}\Big)^9, \nonumber \\
g(k)&\simeq &\frac{4\lambda ^3 (k\lambda)}{21} \Big(\frac{B^2}{2\Gamma (\frac{5}{2})} \Big) \Big(\frac{\mathcal{B}^2}{2\Gamma (\frac{7}{2})} \Big) \Big(\frac{\lambda}{\lambda _d}\Big)^7 ,
\end{eqnarray}
where $ \lambda $ is the length scale on which coherent magnetic fields exist, and $ \lambda _d $ denotes the dissipation scale below which the magnetic power spectrum is exponentially suppressed. Moreover, in the maximally helical case, $ \mathcal{B}^2\sim (\lambda_d/\lambda)B^2 $ \cite{Caprini:2009pr}.
\subsection{Baryon asymmetry calculations}
 Now we can write the GW equation of motion for the right-handed component as 
\begin{equation}
h''_{R}+\frac{2}{u}h'_{R}+h_{R}=\frac{a^2}{k^2 M_p^2}\sqrt{\frac{f(k)-g(k)}{3}},
\end{equation}
where the derivative is taken with respect to the new variable $ u=k\tau $. Also, for the left-handed component the same equation holds with $ \sqrt{(f(k)+g(k))/3} $ as the source. The general solution can be obtained by the following Wronskian method 
\begin{equation}
h_R=c_1(u)\frac{\sin(u)}{u}+c_2(u)\frac{\cos(u)}{u}.
\end{equation}
The second term diverges at small $ u $, hence we consider the first term as the relevant solution for the GWs. We find $ c_1(u) $ as
\begin{eqnarray}
c_1(u)&=&-\int^{u}_{u_i}du' \frac{a^2}{W(u')k^2 M_p^2}\sqrt{\frac{f(k)-g(k)}{3}}\frac{\cos(u')}{u'}\nonumber \\
&=&\int^{u_i}_{u}du' ~\frac{H_0^2\Omega _{rad}~ u'^3 \cos(u')}{k^4 M_p^2}\sqrt{\frac{f(k)-g(k)}{3}}\simeq \frac{H_0^2\Omega _{rad}~ u_i^3}{k^4 M_p^2}\sqrt{\frac{f(k)-g(k)}{3}}. \nonumber \\
\end{eqnarray}
To obtain the second line, the Wronskian determinant is calculated to be $ W(u)=1/u^2 $, and we have kept only the largest part of $ c_1(u) $ which gives the dominant contribution to the final result for the matter asymmetry. Therefore, the GW solutions becomes
\begin{equation}\label{sol}
h_{R, L}=\frac{H_0^2\Omega _{rad}~ \tau _i^3}{k M_p^2}\sqrt{\frac{f(k)\mp g(k)}{3}}\frac{\sin(k\tau)}{k\tau},
\end{equation}
where $ \tau _i\simeq 10^4~\mathrm{s}\simeq 10^{29}~\mathrm{GeV}^{-1} $ is the conformal time at the EWPT. One can define the GW components in terms of the creation and annihilation operators as
\begin{equation}
\hat{h}_R(\tau , \mathbf{x})=\int \frac{d^3k}{(2\pi)^{3/2}}\Big(h_R(\tau , \mathbf{k})\hat{a}_{\mathbf{k}}+h_L^*(\tau , \mathbf{-k})\hat{b}_{\mathbf{-k}}^{\dag} \Big) e^{i\mathbf{k.x}},
\end{equation}
and an analogous relation for $ \hat{h}_L(\tau , \mathbf{x}) $. From this equation, $ \langle R\tilde{R}\rangle $ is given by \cite{Maleknejad:2014wsa}
\begin{equation}\label{rr}
 \langle R\tilde{R}\rangle =\frac{2}{\pi ^2a^4}\int k^3dk\frac{d}{d\tau}\Big(\dot{h}_R(\tau , k)\dot{h}_R^*(\tau , k)-k^2h_R(\tau , k)h_R^*(\tau , k)-R\leftrightarrow L \Big). 
\end{equation}
Since there is an UV cutoff in the helical magnetic field power spectrum, the integral also runs over all physical momentum space up to the UV cut-off, $ k_d=1/\lambda _d $. To find the lepton number density, $ n\equiv L/(\int a^3d^3x) $, we should integrate over the time interval of the EWPT. As can be seen clearly from Eqs.\ (\ref{fg}, \ref{sol}, \ref{rr}), non-zero $ \langle R\tilde{R}\rangle $ is produced by chiral GW components sourced by helical magnetic fields. Furthermore, to gain a net lepton number, CP violating processes should exist. The presence of helical magnetic fields with non-vanishing helicity induces a CP violating source in the system. Besides, the magnetic field affects the scattering of fermions from bubble walls \cite{Ayala:2002xx} which provides an additional source of CP violation and hence the required amount of CP violation can be fulfilled in the system. Putting Eq.\ (\ref{sol}) into Eq.\ (\ref{rr}), we find
\begin{equation}\label{rrr}
\langle R\tilde{R}\rangle \simeq\frac{8H_0^4\Omega _{rad}^2~\tau _i^6}{3\pi ^2a^4M_p^4\tau ^2}\int dk~  k^2g(k)\sin(2k\tau) \simeq \frac{H_0^4\Omega _{rad}^2~\tau _i^6 \lambda B^4}{69\pi ^2a^4M_p^4} \Big(\frac{\lambda}{\lambda _d}\Big)^{9}~\frac{\cos(2k_d\tau)}{\tau ^3}.
\end{equation}
Note that only the helical part, $ g(k) $, contributes to $  \langle R\tilde{R}\rangle $ and incidentally, the dominant term is taken in each step of calculation. This term generates a net lepton number which is subsequently converted to a net baryon number via sphaleron processes. The produced baryon asymmetry is preserved, provided that the EWPT is first-order and so sphaleron processes are suppressed in the broken phase. Taking into account some necessary conditions, including hypercharge neutrality and Yukawa interactions, the ratio of the baryon number to the lepton number can be calculated as $ n_B/n_L \simeq 0.3 $ \cite{Davidson:2008bu}. Using this ratio and integrating Eq.\ (\ref{rrr}) over time, the net baryon number density is obtained as
\begin{equation}\label{20}
n_B=\frac{3H_0^4\Omega _{rad}^2~\tau _i^6 \lambda B^4}{3680\pi ^4M_p^4}\Big(\frac{\lambda}{\lambda _d}\Big)^{9} \int_{H^{-1}}^{1.01~H^{-1}}d\tau \frac{\cos(2k_d\tau)}{\tau ^3},
\end{equation}
where we take $ N=3 $ and the duration of the phase transition is $ 0.01 H^{-1} $ \cite{Kosowsky:1992rz} in which $ H\simeq T^2/M_p =10^{-14}~\mathrm{GeV} $, and $ T\simeq 100~\mathrm{GeV} $ at the EWPT. Finally, to compute the baryon asymmetry, we need to divide $ n_B $ by the entropy density, $ s=2\pi ^2 g_*T^3/45 $, where $ g_* $ is the effective number of massless degrees of freedom which at the EWPT is $ g_*=106.75 $. The dissipation length to correlation length ratio is of order of $ (\lambda _d/\lambda)=10^{-10} $ at the EWPT \cite{Caprini:2009pr}. Notice that $ 1 /\lambda _d $ provides a natural cutoff for the otherwise divergent integral in Eq.\ (\ref{rrr}). Subsequently, the factor $ (\lambda /\lambda _d)^9 $ counteracts the usual suppression factor for processes involving GWs, i.e. the factor $ M_p^{4} $ in the denominator. The remaining factors in Eq.\ (\ref{20}) work out, as we shall see below, to yield reasonable values for $ B $ and $ \lambda $. We now solve the integral numerically. To obtain the baryon asymmetry consistent with observations, $ B $ and $ \lambda $ should be properly determined. For helical magnetic field with inverse cascade evolution, these two parameters can be related to each other. Inverse cascade evolution allows energy shift from small to large scale and will stretch the correlation length. Indeed, cosmological magnetic fields undergone inverse cascade are interesting proposals for the large-scale magnetic fields observed in the galaxies and intergalactic spaces \cite{Ade:2015cva}. We can relate the magnetic field magnitude and its correlation length scale to their present values through the following evolution relations \cite{Fujita:2016igl}
\begin{eqnarray}\label{bla}
B &\simeq &9.3\times 10^{19}\mathrm{G}~ \Big(\frac{T}{10^2~\mathrm{GeV}} \Big)^{7/3}\Big(\frac{B_0}{10^{-14}\mathrm{G}} \Big)^{2/3}\Big(\frac{\lambda _0}{1\mathrm{pc}} \Big)^{1/3}\mathcal{G}_B(T), \nonumber \\ 
\lambda &\simeq &2.4\times 10^{-29}\mathrm{Mpc} \Big(\frac{T}{10^2\mathrm{GeV}} \Big)^{-5/3}\Big(\frac{B_0}{10^{-14}\mathrm{G}} \Big)^{2/3}\Big(\frac{\lambda _0}{1\mathrm{pc}} \Big)^{1/3}\mathcal{G}_{\lambda}(T) 
\end{eqnarray}
where $ \mathcal{G}_B (T) $ and $ \mathcal{G}_{\lambda} (T) $ are $ \mathcal{O}(1) $ factors at the EWPT. Using Eq.\ (\ref{bla}), $ \lambda $ can be obtained in terms of $ B $. Finally, putting helical magnetic field of the order of $ B\simeq 10^4 ~\mathrm{GeV}^2 $ corresponding to $ \lambda \simeq 10^{-3} ~\mathrm{m}\simeq 10^{13} ~\mathrm{GeV}^{-1} $, the observed value of baryon asymmetry can be obtained. Moreover, according to Eq.\ (\ref{bla}) the present values of the required quantities are $ B_0\simeq 10^{-10}~\mathrm{G} $ and $ \lambda _0\simeq \mathrm{10~ kpc} $ which are in agreement with observed large-scale magnetic fields.
\section{Conclusion}
We have presented a new mechanism for EW baryogenesis which relies on the gravitational anomaly sourced by chiral GWs. We assume the existence of a first-order EWPT and helical magnetic fields generating chiral GWs. We solve the GW equation during the PT, and find the gravitational anomaly violating the lepton number. The leptonic number can be transformed to the baryonic number by sphaleron processes. Furthermore, the magnetic helicity is a CP-odd quantity which provides the Sakharov's  second condition and the net baryon number. Thus, it is interesting to note that in our work three Sakharov's criteria are dependent. The baryon asymmetry produced can be compatible with the observed value if the magnetic field and its correlation length scale are of the order of $ B\simeq 10^{24}~\mathrm{G} $ and $ \lambda \simeq 10^{-3} ~\mathrm{m} $, respectively, at the EW scale. Using an inverse cascade evolution, these magnetic fields can be considered as a primordial source for the observed large-scale magnetic fields. Moreover, another important advantage of this idea is that it is not constrained to any specific model and its necessary ingredients might be found in a wide variety of models for electroweak physics.  \\\\

\textbf{Acknowledgments}\\

HA and MA would like to thank M. M. Sheikh-Jabbari for useful discussions. We would like to thank the research council of the Shahid Beheshti University for financial support.


\end{document}